\documentclass[twocolumn,preprintnumbers,amsmath,amssymb]{revtex4}
\usepackage{graphicx}% Include figure files
\usepackage{dcolumn}% Align table columns on decimal point
\usepackage{bm}% bold math

\begin{document}

\title{Quantum Key Distribution with Qubits Encoded in Qutrit}

%\title{Manuscript Title:\\with Forced Linebreak}% Force line breaks with \\

%  \small{R. Kumar,$^1$ R Demkowicz-Dobrza\'{n}ski,$^1$ and K. Banaszek$^{1,2}$}\\
% \textit{ \small{$^1$Faculty of Physics, University of Warsaw, ul.Ho\.{z}a 69,
%  00-681 Warsaw, Poland}\\
%  \small{$^2$Institute of Physics, Nicolaus Copernicus University, ul.\ Grudzi\c{a}dzka 5, 87-100 Toru\'{n}, Poland}}
%\author{R.Kumar}
% \altaffiliation{Faculty of Physics, University of Warsaw, ul.Ho\.{z}a 69,
%  00-681 Warsaw, Poland.}
%\author{R. Demkowicz-Dobrza\'{n}sk}
%\altaffiliation{Faculty of Physics, University of Warsaw,
%ul.Ho\.{z}a 69,  00-681 Warsaw, Poland.}
%\author{K. Banaszek}
%\altaffiliation{Faculty of Physics, University of Warsaw,
%ul.Ho\.{z}a 69,  00-681 Warsaw, Poland.}

%\author{R.Kumar$^{1}$, R.Demkowicz-Dobrza\'{n}ski$^{1}$, K.Banaszek$^{1,2}$\\
%\small $^{1}$Faculty of Physics, University of Warsaw, ul.Ho\.{z}a
%69, 00-681 Warsaw, Poland\\
%$^{2}$Institute of Physics, Nicolaus Copernicus University, ul.\
%Grudzi\c{a}dzka 5, 87-100 Toru\'{n}, Poland}
%
%\email{rupesh.k@fuw.edu.pl}

\author{R.Kumar$^{1}$}
 \email{rupesh.k@fuw.edu.pl}
\author{R.Demkowicz-Dobrza\'{n}ski$^{1}$}
\author{K.Banaszek$^{1,2}$}
 \affiliation{$^{1}$Faculty of Physics, University of Warsaw, ul.Ho\.{z}a
69, 00-681 Warsaw, Poland. \\$^{2}$Institute of Physics, Nicolaus
Copernicus University, ul.Grudzi\c{a}dzka 5, 87-100 Toru\'{n},
Poland}

%\begin{center}
%  \large{\textbf{Quantum Key Distribution with Qubits Encoded in Qutrit.}}\\
%  \small{Rupesh Kumar,$^1$ Ludmi{\l}a Praxmeyer,$^2$ Rafa{\l} Demkowicz-Dobrza\'{n}ski,$^1$ and Konrad Banaszek$^{1,2}$}\\
% \textit{ \small{$^1$Faculty of Physics, University of Warsaw, ul.Ho\.{z}a 69,
%  00-681 Warsaw, Poland}\\
%  \small{$^2$Institute of Physics, Nicolaus Copernicus University, ul.\ Grudzi\c{a}dzka 5, 87-100 Toru\'{n}, Poland}}
%  \end{center}
%  %\begin{abstract}

\begin{abstract}
We present a novel one-way quantum key distribution protocol based
on 3-dimensional quantum state, a qutrit, that encodes two qubits
in its 2-dimensional subspaces. The qubits hold the classical bit
information that has to be shared between the legitimate users.
Alice sends such a qutrit to Bob where he decodes  one of the
qubit and measures it along the random Pauli basis. This scheme
has higher secure key rate at longer transmission distance than
the standard BB84 protocol.
\end{abstract}

\maketitle

Quantum key distribution (QKD)  allows two legitimate users,
namely Alice and Bob, to share secure key for cryptographic
purpose. The security of the shared key is guaranteed by the laws
of quantum physics. Besides its unconditional security, a
promising QKD protocol needs to meet higher
 key generation rate as well as longer transmission distance. Since
its first proposal in 1984  the protocol BB84~\cite{BB84} has been
scrutinized and proven  secure under a vast range of eavesdropping
strategies
~\cite{ShorPreskill,GisinReview2002,ScaraniReview2009,collectiveattack,Lutkenhaus1996}.
There are also other protocols ~\cite{SARG04,DECOY,DPSQKD,COW},
merited by their own account, that enhance the secure key
generation rate and transmission distance beyond the BB84
protocol.

However, efforts are also made on BB84 protocol for increasing its
throughput. For example, decoy state BB84~\cite{DECOY} increases
the transmission distance by securing it against the photon number
splitting (PNS) attack. Using integrated optics based on planar
lightwave circuit technology~\cite{NAMBU} increases the key
generation rate of  BB84 protocol implemented with time-bin
qubits. Along in this direction, we propose a one-way scheme that
utilizes 3-dimensional quantum state, a qutrit, for performing
QKD. Performance of qutrit based protocols have been studied
before in ~\cite{QUTRIT1, QUTRIT2, QUTRIT3, QUTRIT4}.
 The main feature of our scheme is that the qutrit encodes
two BB84-qubits in its 2-dimensional subspaces. By saying
 BB84-qubits we address the qubits originally prepared along the random Pauli basis  for
 performing the protocol BB84. In order to extract the classical
 bit information one has to first decode the qubits from the qutrit and
 then measure it along the correct basis.

 The advantage of this scheme is that Eve, also Bob, cannot
 decode the qubits  with certainty. As a
 result,  eavesdropping strategies assisted with quantum memory~\cite{PNSattack} may
 not be fully effective for stealing complete bit of information.
 This implies that the practical implementation of our scheme using weak coherent source
 can have comparatively higher mean photon number, $\mu$, and higher key rate at
 longer transmission distance than the  standard BB84 scheme. Additionally, using single photon
 pulses may drain less information to Eve during individual attacks~\cite{LutkenhausIndividualAttack} .

In the following, we will describe  our scheme by featuring its
 encoding and decoding procedure by adopting the technique
given in ~\cite{Grudka}. This will follow  a comparison of the
performance of the proposed protocol with standard BB84
 protocol,  while they are under collective attacks.

\begin{figure}
  \begin{center}
   \includegraphics[width=0.5\textwidth]{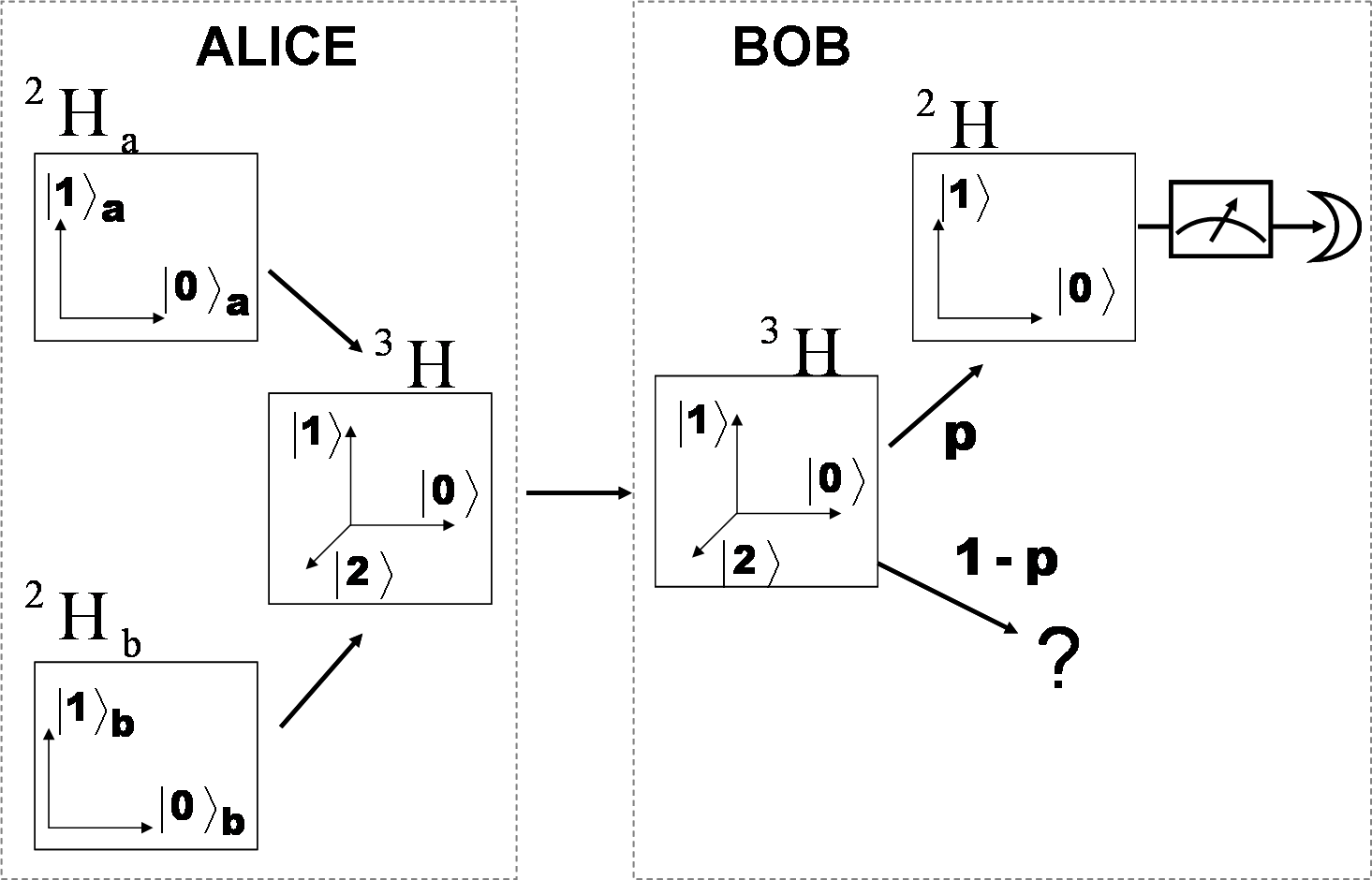}
  \end{center}
  \caption{The scheme for qkd with qubits encoded in qutrit.
  Alice encodes two BB84 qubits that span the Hilbert space $^{2}{H}_{a}$ and $^{2}{H}_{b}$
 in a qutrit in $^{3}{H}$ and sends to Bob.
  He decodes one of the qubit from the qutrit with a probability $p$ and measure the qubit along Pauli Z or X basis.
  With a probability $1-p$ he fails to decode the qubit and the protocol fails.}
  \label{fig:QUTRITprotocol}
\end{figure}

 Let us start the description of our
protocol, see Fig.~\ref{fig:QUTRITprotocol}, by considering two
qubit states $|\psi_{a}\rangle,
(|0\rangle_{a}+e^{i\varphi_{a}}|1\rangle_{a})/\sqrt{2}$ and
$|\psi_{b}\rangle= (|0\rangle_{b}+e^{i \varphi_{b}
}|1\rangle_{b})/\sqrt{2}$, that span the 2-dimensional Hilbert
spaces $^{2}{H}_{a}$ and ${^{2}{H}}_{b}$, respectively. The basis
of the joint state $|\psi\rangle=|\psi_{a}\rangle \otimes
|\psi_{b}\rangle$ are labeled as: $|i\rangle_{a}|j\rangle_{b}\to |
|3j-i| \rangle$. Let
$|\phi\rangle=(|0\rangle+|1\rangle+|2\rangle)/\sqrt{3}$ be a
qutrit state in $^{3}{H}$. It is possible to encode the two qubits
in the qutrit  by performing a projective measurement on the joint
state $|\phi\rangle \otimes |\psi\rangle $ using the projector
\begin{equation}
\Pi =\sum_{i=0}^{2}|i\rangle\langle i|\otimes| i\rangle\langle i|.
\label{ALICEProjector}
\end{equation}
The resultant qutrit state, after the unitary operation
$|i\rangle\otimes|j\rangle \to |i\rangle\otimes|j-i\rangle$, is
\begin{equation}
|\Phi\rangle= \frac{1}{\sqrt{3}}(|0\rangle+e^{i
\varphi_{a}}|1\rangle+e^{i( \varphi_{a}+\varphi_{b})}|2 \rangle).
\label{ALICEqutrit}
\end{equation}

By projecting the qutrit $|\Phi\rangle$ onto  a 2-dimensional
subspace the encoded qubits can be retrieved with  probability
2/3. The respective projectors are:
\begin{equation}
\Pi_{1} = |0\rangle\langle0|+|1\rangle\langle1|\hspace{2 mm}
\textit{and} \hspace{2 mm}
 \Pi_{2} =
|1\rangle\langle1|+|2\rangle\langle2|.
 \label{ProjBob}
\end{equation}
And the resultant qubit states are,
\begin{equation}
  |\psi\rangle_{i} =
\frac{\Pi_{i}|\Phi\rangle}{\sqrt{\langle\Phi|\Pi_{i}|\Phi\rangle}}=
\left\{
  \begin{array}{l l}
    \frac{1}{\sqrt{2}}(|0\rangle + e^{i \varphi_{a}} |1\rangle) & : \text{$i$=1}\\
    \frac{1}{\sqrt{2}}(|1\rangle + e^{i \varphi_{b}}
|2\rangle) & : \text{$i$=2}\\
  \end{array} \right.
\end{equation}
In the above equation global phases are ignored.

Now, consider Alice prepares two qubits in one of the eigen states
of Pauli Z and X operators, by randomly choosing the value of
$\varphi_{a}$ and $\varphi_{b}$ from the set
$S=\{0,\frac{\pi}{2},\pi,\frac{3\pi}{2}\}$. She then sends them to
Bob one after other where he measures them, randomly, along Pauli
Z or X basis.  This is the standard BB84 protocol, executing twice
for the distribution of secret key between the legitimate users.
Here, we would like to examine how the qutrit state results from
the encoding process can  be used for performing such a secret key
distribution if Alice would have prepared the initial qubits in
eigen states of Pauli Z and X operators. We call this scheme as
`Qutrit' QKD, hereafter. An experimental demonstration of encoding
and decoding procedure is given in ~\cite{Dusek}. However, For the
simplicity and thence practicality, we can assume that a qutrit
state prepared in Eq.~\eqref{ALICEqutrit} inherently encodes
qubits in its subspace. This assumption helps Alice to starts the
protocol straight from the qutrit $|\Phi\rangle$ by obviating the
encoding process.  Alice chooses the values of $\varphi_{a}$ and
$\varphi_{b}$ from the set $S$ such that qubits,
$|\psi_{a}\rangle$ and $|\psi_{b}\rangle$,
 in the subspaces are in Pauli eigen states. She then sends the
 qutrit to Bob through a quantum channel. Upon
receiving the qutrit, Bob  randomly selects the projectors
$\Pi_{1}$ or $\Pi_{2}$  and decodes one of the qubits. He then
measures the qubit along the random Pauli Z or X basis and
registers the measurement outcome together with the order (say,
first or second) of the qubit. This creates his raw key with
probability 2/3. Bob fails to decode the qubit with probability
1/3 and the protocol fails.
%////quantumcommunication ends, from
%the clicks make key... explain it.
During key sifting process, Bob announces which qubit he has
decoded  and on which basis he has measured it. If the basis
matches for the respective qubit then Alice agrees to keep the key
bit for further process such as error correction and privacy
amplification. Otherwise, they discard it. This procedure exactly
follows the classical part of BB84. Since our proposed qutrit
protocol shares many features of the BB84 protocol,  we would like
to compare the performance of both of the protocols in terms of
the secure key rate.

For the practical reason we consider the protocols are implemented
using weak coherent source. Such a source can easily be realized
using only standard semiconductor laser and calibrated
attenuators. The photon number statistics of a weak coherent
source follows Poisson distribution. Therefore, the probability of
finding $n$ photons  in the quantum state generated by Alice is
$P(n,\mu)= \mu^{n}e^{-\mu}/n!$, where  $\mu$ is the mean photon
number per signal. Accordingly, with certain  probability, a weak
coherent source emits signals that contain more than one photon.
%Accordingly, the probability that a signal contains two photon is
%$\approx \mu/2$.
Such multi-photon signals breach the security of the QKD protocol
because  an eavesdropper gains partial or complete information on
the shared key without revealing her presence to the legitimate
users. Probability of multi photon pulses can be made arbitrarily
small by choosing $\mu<1$. But  very low value of $\mu$ results in
overall reduction in signal generation rate. Therefore, there is a
trade-off between the value of $\mu$ chosen by Alice and  secure
key rate of the protocols, for each transmission distance.  For
example, at higher transmission distance, Eve may replace the
lossy quantum channel with a lossless channel and performs PNS
attack on the multi-photon pulses. She takes advantage on the
attenuation of the lossy channel by blocking as much as single
photon pulses such that the total signal detected by bob remains
constant. Therefore, for transmission thorough a lossy channel,
value of $\mu$ must be chosen smaller than that of an ideal
lossless channel.
%It has been shown that the value of  $\mu$
%varies as the of the channel[REF].
We can see that the proposed qutrit protocol allows Alice to
choose comparatively higher $\mu$ than the BB84 protocol and that
in turn considerably enhance the key generation rate at higher
transmission distances.
 %This has  straight implication in the eavesdropping
%strategy adopted by Eve- different attacks on single and multi
%photon signals.

In the following  we compare the performance of the BB84 and
Qutrit protocols by quantifying the achievable secret key rate $K$
while considering the protocols are under collective
attacks~\cite{collectiveattack}. During this attack, Eve
independently  attacks each quantum signal sent by Alice. She can
store her ancillary states in quantum memory and  wait for the
classical post-processing and adopt a best measurement strategy to
extract the information. Following an approach similar to that in
~\cite{ScaraniReview2009} we will now consider the security aspect
of the protocol while they are under collective attack. The secret
key rate of a QKD protocol can be defined as

\begin{equation}
K = P_{accept}P_{sift}R_{raw}[1-h(Q)-I_{E})].
 \label{SecretKey}
\end{equation}

In the above equation, $P_{accept}$ depends on the protocol
implementation and is the probability that a detector click
contributes to the raw key, $P_{sift}=1/2$ accounts for the
fraction of raw key discarded due to wrong measurement basis,
$R_{raw}=(R_{sig}+2p_{d}(1-R_{sig}))$ is the raw key rate in which
$R_{sig}=1-e^{-\mu\Gamma_{q} \Gamma_{b} \eta} \approx  \mu
\Gamma_{q} \Gamma_{b} \eta$  is the total detector clicks due to
photons that survived the attenuation $\Gamma_{q}$ of the quantum
channel and $\Gamma_{b}$ of Bob's apparatus with detectors of
average efficiency $\eta$ and dark count probability $p_{d}$. The
above approximation holds for the mean photon number per pulse
$\mu<1$. The channel attenuation related to transmission distance
$l$ as $\Gamma_{q}=10^{-\alpha l/10}$, where $\alpha$ is the
attenuation coefficient. $I_{E}$ is the  amount  of information
eavesdropped by Eve. $h(Q)$is the amount of information leaked to
Eve during error correction procedure and $h(.)$is the Shannon
entropy. The qber $Q$ is a function of $\mu$ and is defined as
\begin{equation}
Q=p_{d}(1-R_{sig})/R_{raw}+Q_{opt}.
\label{QBER}
\end{equation}

The first term in the above equation is the error rate due to dark
counts of Bob's detectors. The constant $ Q_{opt}$ accounts for
the errors from optical misalignment of Bob's apparatus. We
attribute all sources of error to Eve.

 During collective attacks, Eve learns the
number of photon $n$ present in the weak coherent pulse sent by
Alice and adopt a best attacking strategy that  maximize her
information on the final secure key shared by the users. On single
photon pulses, Eve can gain information at the expenses of
introducing as error $\varepsilon_{1}$.  The amount of information
she obtains on single photon pulse is $I_{E,1}=h(\varepsilon_{1})$
~\cite{ScaraniReview2009}.   For example, she can perform simple
intercept and resend attack that creates an error with probability
1/4 and gain 1/2 a bit of information. For multi-photon pulses,
$n\geq2$, the optimal attack is PNS attack during which Eve
forwards one photon to Bob and keeps the rest in her quantum
memory. She makes no error and gain complete bit of information,
i.e., $\varepsilon_{n\geq2}=0$  and $I_{E,n\geq2} = 1$. However,
Eve learns  zero information while Alice sends vacuum pulse and
Bob gets detection event due to detector dark counts. Let us
consider the parameter $Y_{n} = R_{n}/R_{raw}$ be the probability
that Bob gets a sifted key from $n$-photon pulse sent by Alice
with probability $P_{A}(n)$ . Here, $R_{n}= P_{sift}P_{A}(n)f_{n}$
and $f_{n}$ is the probability that Eve forwards a single photon
to Bob for $n$-photon pulse. Eve can optimize the value of $f_{n}$
for the total detection rate $R_{raw}$. The overall information
exposed to Eve is
\begin{eqnarray}
 I_{E}&=&\max\left[Y_{1}h(\varepsilon_{1})+(1-Y_{0}-Y_{1})\right]\nonumber\\
 & = &1-\min\left\{Y_{0}+ Y_{1}[1+h(\varepsilon_{1})]\right\}. \label{EveInfo}
 \end{eqnarray}
In order to maximizing  her information, Eve's optimal attack
should be compatible with the measured parameters available to the
legitimate users: i.e., key rate $R_{raw}$ and the qber
$Q=Y_{1}\varepsilon_{1}$. This is achieved by minimizing $Y_{1}$
in Eq.~\eqref{EveInfo} by setting $f_{0}=0$ and $f_{n\geq2}=1$.
Therefore, we can rewrite Eq.~\eqref{EveInfo} for the protocol
BB84 as
\begin{equation}
I_{_{E}}^{^{BB84}} = 1-Y_{1}[1-h(\varepsilon_{1})],
 \label{Y1BB84Final}
\end{equation}
where,
\begin{equation}
Y_{1} = 1-Y_{0}-Y_{n\geq2}=1-P_{sift}P_{A}(n\geq2)/R_{raw}.
 \label{Y1}
\end{equation}

%\begin{equation}
%I_{E} = p_{A}(n\geq2)/R_{raw}[h(R_{raw}Q/1-p_{A}(n\geq2)-1].
% \label{Y1BB84Final}
%\end{equation}

The corresponding achievable secret key rate for the protocol BB84
is
\begin{equation}
K_{_{BB84}} = R_{raw}\{Y_{1}[1-h(\varepsilon_{1})]-h(Q)\}/2.
 \label{SecretKeyBB84}
\end{equation}

Here we have taken $P_{accept}=1$. It is true for the
implementation with polarization coding of the BB84 protocol.
However, we can take the same value for the phase encoding scheme
as well~\cite{Townsend1994}.

Now, let us consider information eavesdropped on qutrit protocol
during collective attack. An attack against qutrit protocol said
to be successful only when both Eve and Bob decode the qubit from
the same subspace. On single photon qutrit pulse Eve decodes one
of the qubits with probability 2/3 and  performs her measurement.
She then forwards a new qutrit to Bob that encodes two new qubits
in its subspaces. The quantum state of the qubit in the subspace
under attack corresponding  to Eve's  measurement result. On the
other hand, since Eve has no information on the state of the qubit
which was not under her attack, she prepares the other qubit in
random Pauli basis. On failure of decoding, she sends a vacuum
pulse to Bob. One may argue that empty pulses reduce Bob's overall
detection rate but it is not the case: since Eve can optimize  the
function $f_{n}$ compatible with Bob's detection rate. Therefore,
the probability that Eve and Bob decodes the qubit from the same
subspace becomes 1/2. For each successful decoding, Eve gets
$I_{E,1}=h(\varepsilon_{1})$ bits of information. Here, as
mentioned earlier, $\varepsilon_{1}$ is the error introduced by
Eve. On the other hand, if Bob decodes the qubit from the subspace
which was not under attack, not only Eve gets zero information but
also she introduce an error with probability 1/2. Therefore, on
average, Eve gets $h(\varepsilon_{1})/2$ bits of information and
creates $(\varepsilon_{1}+1/2)/2$ bits of error.
%
% There is another strategy at Eve's
%disposal that gives comparatively higher information at lower
%error probability: send a qubit to Bob corresponding to the
%decoded subspace rather than a qutrit. A projector on this qubit
%onto the other subspace give a random detection outcomes with
%probability 1/3 out of which half of the detection  are counted as
%error. With a probability 1/3 he gets no detection and considered
%as loss. In effect, Eve gets $h(\varepsilon_{1})2/3$ bits of
%information and creates $\varepsilon_{1}2/3+1/6$ bits of error.

 There is another strategy at Eve's
disposal: send a qubit to Bob corresponding to the decoded
subspace rather than a qutrit. This lowers Bob's detection
probability in the second subspace to 1/2 out of which half of the
detections contribute to errors. Bob attributes the reduction in
the detection probability to the channel loss though, Eve
compensates it by tuning the transfer function $f_{n}$. On
average, Eve gains $2h(\varepsilon_{1})/3$ bits of information at
the expense of $2\varepsilon_{1}/3+1/6$ bits of error. It can be
seen that, this strategy gives comparatively higher information
gain to Eve at lower error probability. Therefore the optimal
attacking  strategy for Eve on single photon pulse is forward
qubit rather than qutrit, to Bob.

% It
%must be noted that this qubit creates a random detection outcome,
%with probability 1/3, at Bob's measurement device is half of his
%decoding probability and he gets no
%
%On the other hand, if Bob opts to decode the other qubit,
%regardless of his measurement basis, he gets random measurement
%outcome with probability 1/3 and half of this probability he gets
%errornous
%
%On the other hand, if the subspace mismatches, she gets zero
%knowledge and introduce an error with probability 1/2. However, if
%Bob tries to decode the qubit from the other subspace, he gets
%random measurement outcome with probability 1/3
%
% This
%can be made clear by considering an example. TELL ABOUT bob's
%detectin with probability 1/2. 1/2 of the detection gives error.
%
%other 1/2bob can not detect. which he will consider as loss. to
%compensate this lose eve can control the function fn. so on
%average eve gets 2/3 h epsilon info and make error 2/3E+1/6
%
% eve
%gets $h(\varepsilon_{1})2/3$ error $(\varepsilon_{1}2/3) + 1/6 $
%It must be noted that this is half of the
%
% On average,
%Eve gets $h(\varepsilon_{1})1/2$ bits of information and creates
%$(4\varepsilon_{1}+1)/4$ error. Therefore the optimal strategy for
%Eve is not to send qubit but qutrit.

 PNS attack on multi-photon
qutrit pulses is described as follows. After the basis revelation
by the users, Eve projects the qutrit  onto exactly the same
subspace from which Bob had decoded his qubit. This happens with a
probability 2/3 and she measures the decoded qubits along the
correct basis and gains full information. Therefore, for
two-photon pulses Eve gets 2/3 bit of information, on contrary to
one bit of information in PNS attack against BB84 protocol. For
three-photon pulses, Eve gets 8/9 bit of information which can be
approximated to 1 bit of information. Moreover, at very low value
of mean photon number the contribution of three-photon pulses are
negligible.  By setting the values of $f_{0}=0$ and $f_{n\geq2}=1$
for maximizing her information and the qber
$Q=Y_{1}(2\varepsilon_{1}/3+1/6)$, we can rewrite
Eq.~\eqref{EveInfo} for the qutrit protocol as
%\begin{equation}
%I_{_{E}}^{qutrit} = 1-Y_{1}[1-2h(Q/Y_{1})/3]-Y_{2}/3,
% \label{Y1BB84Final}
%\end{equation}
%
%\begin{equation}
%I_{_{E}}^{qutrit} = 1-Y_{1}[1-2h(3(Q/Y_{1}-1/6)/2)/3]-Y_{2}/3,
% \label{Y1BB84Final}
%\end{equation}

\begin{equation}
I_{_{E}}^{qutrit} = 1-Y_{1}[1-2h(\varepsilon_{1})/3]-Y_{2}/3,
 \label{Y1BB84Final}
\end{equation}

where, $Y_{1}$ is same as that defined in Eq.~\eqref{Y1} and
\begin{equation}
Y_{2} = P_{sift}P_{A}(n=2)/R_{raw}.
 \label{Y2}
\end{equation}

Finally, the secure key rate of the qutrit protocol is
\begin{equation}
K_{qutrit} =
R_{raw}\{Y_{1}[1-2h(\varepsilon_{1})/3]+Y_{2}/3-h(Q)\}/3,
 \label{SecretKeyQUTRIT}
\end{equation}

with  $P_{accept}$=2/3 as the probability of successful decoding
of qubit from the qutrit.

\begin{figure}
  \begin{center}
   \includegraphics[width=0.5\textwidth]{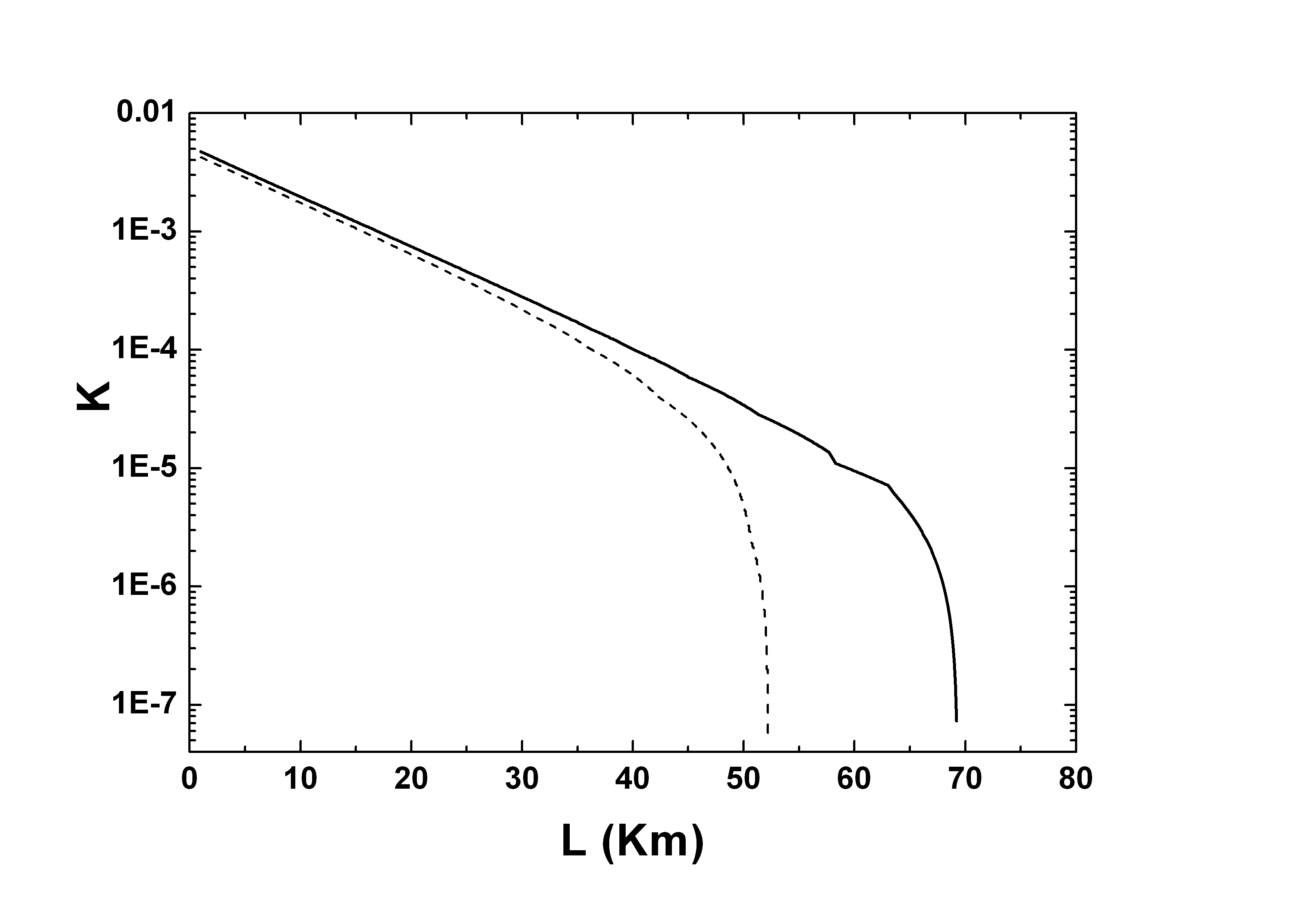}
  \end{center}
  \caption{Key rate K vs transmission distance L. The figure shows the  comparison of Eq.~\eqref{SecretKeyBB84}
  and Eq.~\eqref{SecretKeyQUTRIT} in terms of transmission distance. The solid and dashed lines are for qutrit and  BB84 protocol, respectively.
  The parameters used for this  plot are: $p_{d}$=$10^{-5}$,$\eta$=10$\%$, $Q_{opt}$=0.5$\%$, $\Gamma_{b}$=0.5. The qutrit
protocol gives secure transmission distance of 69 km compare to
52km of BB84 protocol.}
  \label{fig:Comparison}
\end{figure}

The Fig.~\ref{fig:Comparison}. shows the comparison of secure key
rate of BB84 and qutrit protocol in terms of transmission
distance. The mean photon number $\mu$ is numerically optimized
and used in Eq.~\eqref{SecretKeyBB84} and
Eq.~\eqref{SecretKeyQUTRIT} for maximizing the respective secure
key rate for each transmission distance.
%It has been found that
%the average value of optimal $\mu$  is 0.1 for BB84 and  0.17 for
%qutrit protocol, respectively. Higher $\mu$ for qutrit protocol is
%because of lower information drain to Eve from multi-photon
%pulses.
%%We have assumed that the error rates of both of the
%%protocols are identical.
 It  can be seen that, despite of
imperfect qubit decoding probability, the qutrit protocol gives
same key rate as BB84 at shorter distance. More importantly, it
gives comparatively higher key rate at longer distance, which is
the linchpin of the qutrit protocol. Under the same experimental
conditions, maximum secure transmission distance of 69km is
obtained for qutrit protocol which is remarkably higher than 52 km
of BB84 protocol.

In conclusion, we have presented a one-way
 protocol based on qutrit for secure key distribution between two legitimate
 users. The information encoded in the qubit subspaces of the
 qutrit  make the protocol  highly tolerable to
 photon number splitting attack; thanks to
 non-unity qubit decoding probability. Additionally, it is more robust
 against attacks on single photon pulses.
 This is a promising feature and  may encourage  an experimental
realization of  the proposed protocol in the near future.

 We thank  L. Praxmeyer  and M. Lucamarini for fruitful discussions. This
 work is supported by European Commission under the Integrating
 Project Q-ESSENCE.

%%%%\bibliography{apssamp}% Produces the bibliography via BibTeX.

%
%\bibliography{MyBib}
%\bibliographystyle{unsrt}

\end{document}